\documentclass[prl,superscriptaddress,longbibliography,twocolumn]{revtex4}
\usepackage{graphicx}
\usepackage{bm}
\usepackage{url}
\usepackage{tikz}
\usetikzlibrary{calc}
\usepackage{pgfplots}
\usepackage{overpic}

\usepackage{color}
\usepackage[normalem]{ulem}

\newcommand{\bea}{\begin{eqnarray}}
\newcommand{\eea}{\end{eqnarray}}

\newcommand{\be}{\begin{equation}}
\newcommand{\ee}{\end{equation}}
\newcommand{\bk}{{{\bf{k}}}}

\newcommand{\bQ}{{{\bf{Q}}}}

\newcommand{\beal}{\begin{align}}
\newcommand{\eeal}{\end{align}}
\newcommand{\ra}{\rangle}
\newcommand{\la}{\langle}

\newcommand{\dg}{{\dagger}}
\newcommand{\pdg}{{\phantom\dagger}}

\begin{document}

\title{The Curious Case of NiRh$_2$O$_4$: A Spin-Orbit Entangled Diamond Lattice Paramagnet}
\author{Shreya Das}
\affiliation{Department of Condensed Matter Physics and Materials Science, S.N. Bose National Centre for Basic Sciences, Kolkata 700098, India.}
\author{Dhani Nafday}
\affiliation{School of Mathematical and Computational Sciences, Indian Association for the Cultivation of Science, Kolkata 700 032, India.}
\author{Tanusri Saha-Dasgupta}
\affiliation{Department of Condensed Matter Physics and Materials Science, S.N. Bose National Centre for Basic Sciences, Kolkata 700098, India.}
\affiliation{School of Mathematical and Computational Sciences, Indian Association for the Cultivation of Science, Kolkata 700 032, India.}
\author{Arun Paramekanti}
\email{arunp@physics.utoronto.ca}
\affiliation{Department of Physics, University of Toronto, Toronto, Ontario, Canada M5S 1A7.}
\date{\today}

\begin{abstract}
Motivated by the interest in topological quantum paramagnets in candidate spin-$1$ magnets, we investigate the 
diamond lattice compound NiRh$_2$O$_4$ using {\it ab initio} theory and model Hamiltonian approaches.
Our density functional study, taking into account the unquenched orbital degrees of freedom, shows stabilization of  $S\!\!=\!\! 1$,  $L\!\!=\!\! 1$ state. 
We highlight the importance of spin-orbit coupling, in addition to Coulomb correlations, in driving the insulating gap, and uncover frustrating 
large second-neighbor exchange mediated by Ni-Rh covalency. A single-site model Hamiltonian incorporating the large tetragonal distortion is shown to
give rise to a spin-orbit entangled non-magnetic ground state, largely accounting for the entropy, magnetic susceptibility, 
and inelastic neutron scattering results. Incorporating inter-site exchange within a slave-boson theory, we show that 
exchange frustration can suppress exciton condensation. We capture the dispersive gapped 
magnetic modes, uncover ``dark states'' invisible to neutrons, and make predictions.
\end{abstract}

\maketitle

{\it Introduction. --- } Symmetry protected topological phases of quantum matter, e.g.,
two dimensional (2D) and 3D topological insulators \cite{TopoIns_RMP2010,TopoIns_RMP2011}, Weyl semimetals \cite{WeylSM_ARCMP2017}, and
topological superconductors \cite{TopoIns_RMP2011}, have been extensively discussed in the context of electronic systems. Following these
remarkable discoveries, interacting spins and bosons have also been theoretically proposed to support symmetry-protected
topological ground states with conventional bulk excitations but unusual gapless or gapped edge states \cite{Vishwanath_PRX2013,Levin_PRB2012,Pollmann_PRB2012,Levin_PRL2013,Senthil_ARCMP2015}.
Recently, there has been an exciting proposal that certain $S\!\!=\!\!1$ spin models on the diamond lattice may realize a time-reversal symmetry
protected topological 
quantum paramagnet \cite{ChongWang_PRB2015}, a stable 3D analogue of the $S\!\!=\!\!1$ Haldane chain \cite{Haldane_PRL1983,AKLT_PRL1987}.

 This has led to a renewed interest in candidate spinel materials AB$_2$O$_4$ with A-site spins living on the diamond lattice.
Previous studies of A-site magnetic spinels, such as MnSc$_2$S$_4$ ($S \!\!=\!\! 5/2$) and CoAl$_2$O$_4$ ($S\!\!=\!\! 3/2$), revealed degenerate spin 
spirals driven by frustration 
\cite{Loidl_PRL2004,Loidl_PRB2005,Bergman_NPhys2007,Bernier_PRL2008,Ruegg_NPhys2016,Oitmaa_PRB2019}. 
On the other hand, FeSc$_2$S$_4$ shows weak N\'eel order in proximity to a non-magnetic ground state induced by spin-orbit coupling (SOC) 
\cite{GangChen_PRL2009,GangChen_PRB2009,Plumb_PRX2016}.
The search for $S\!\!=\!\! 1$ topological paramagnets has recently led to an intense investigation of
NiRh$_2$O$_4$ using a variety of tools \cite{Chamorro_PRM2018}.

NiRh$_2$O$_4$ is an unusual example of spin-$1$ 3d ions on the tetrahedrally coordinated A site, which
is structurally stabilized by placing 4d Rh$^{3+}$ ion at the octahedrally coordinated B-site. While NiRh$_2$O$_4$ is
cubic at high temperature \cite{Blasse_PL1963,Chamorro_PRM2018}, it transforms into a tetragonal phase below $T \! \sim \! 380$ K. 
Remarkably, in contrast to expectations from a Jahn-Teller
mechanism which would favor $c/a \!<\! 1$ and an $S\!\!=\!\! 1$ ground state with quenched orbital angular momentum,
the tetragonal phase is found to be elongated with $c/a\!  \approx \! 1.05$. 
Such a tetragonal distortion, with $c/a \!\!>\!\! 1$, leaves the $t_2$ states of Ni partially filled,
with orbital degrees of freedom unfrozen, allowing spin-orbit coupling (SOC) to play an important role. 
The mechanism for tetragonal distortion thus relies on SOC-induced orbital ordering, 
as previously discussed \cite{OlegT_PRL2004,Maitra_PRL2007} in the context of the B-site active spinel ZnV$_2$O$_4$.

An early theoretical study \cite{GangChen_PRB2017} of NiRh$_2$O$_4$ considered a model with antiferromagnetic (AFM) first and
second-neighbor Heisenberg exchanges ($J_1$ and $J_2$), applicable to frustrated spinels, and proposed that the 
non-magnetic ground state might arise from large single-ion anisotropy $D S_z^2$, with $D \!\!>\!\! 0$ favoring local
$S_z\!\!=\!\!0$. A pseudospin functional renormalization group study of the $J_1$-$J_2$ model \cite{Buessen_PRL2018}
found that while the $S\!\!=\!\!1$ case favors a quantum
spiral spin liquid ground state, the impact of tetragonal distortion or large $D/J_1\!\! \gtrsim \!\! 8$ is to respectively favor N\'eel order or the 
$S_z\!\!=\!\!0$ ground state. Both studies effectively ignored orbital degrees of freedom.
More recently, it was proposed \cite{GangChen_J0_2018} that strong SOC with a tetrahedral crystal field could support 
a $J_{\rm eff}\!\!=\!\!0$ state at $d^8$ filling, generalizing the idea of $J_{\rm eff}\!\!=\!\!0$ insulators for $d^4$ filling in an 
octahedral crystal field \cite{Khaliullin_J0_PRL2013,Khaliullin_J0_PRB2014,Trivedi_J0_PRB2017}; however, this might be overwhelmed
by other energy scales (e.g., distortions or inter-site exchange) given weak SOC for Ni$^{2+}$. On the experimental front, the inelastic neutron scattering (INS) results
\cite{Chamorro_PRM2018} on NiRh$_2$O$_4$ were analyzed using spin-wave theory of an AFM state despite the absence of N\'eel order.

A satisfactory theoretical description of NiRh$_2$O$_4$ is thus lacking. Here, we combine
first-principles density functional theory (DFT) and a model Hamiltonian study to unravel the curious case of
NiRh$_2$O$_4$, explaining existing data and making predictions for future experiments.

{\it Density functional theory. --- } We have carried out a first-principles study of NiRh$_2$O$_4$ in full-potential
all electron approach of linear augmented plane wave (FLAPW) method \cite{wien2k}, muffin-tin orbital method 
\cite{lmto,nmto}, as well as in pseudo-potential plane wave
basis \cite{vasp} with projected augmented potential (PAW) \cite{paw}. The exchange-correlation functional was chosen to be
generalized gradient approximation (GGA) \cite{gga}, supplemented with onsite Hubbard correction GGA+$U$ \cite{gga+u}.
Calculational details may be found in the Supplementary Material (SM) \cite{sm}.

\begin{figure}[b]
\includegraphics[width=0.48\textwidth,keepaspectratio]{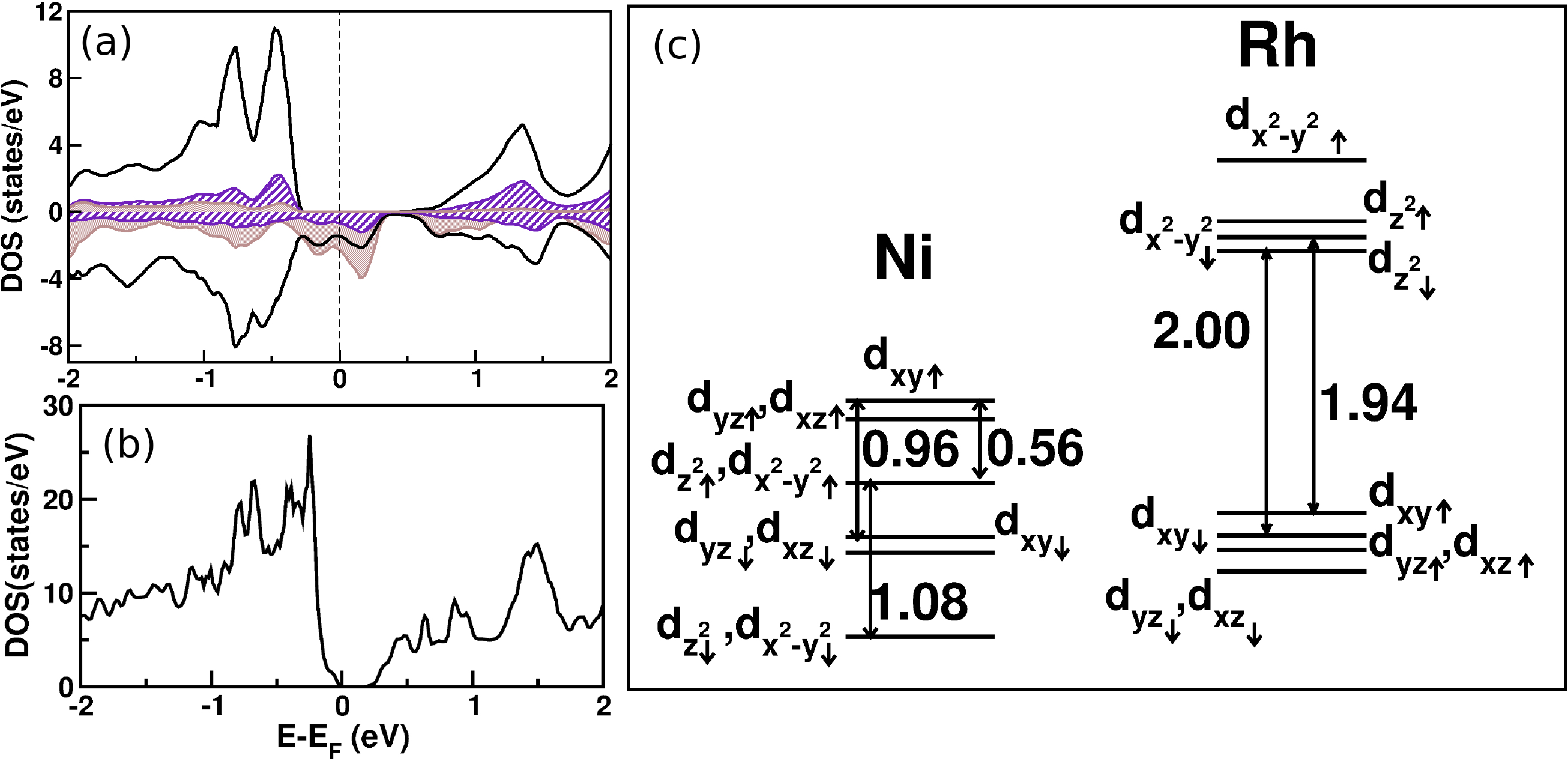}
\caption{(Color online) (a) The GGA+$U$ electronic structure of NiRh$_2$O$_4$ in low-temperature tetragonal phase. States projected onto Ni $d$, Rh $d$ and O $p$ characters are shown as grey-shaded, black-solid line, hatched areas, respectively. (b) The GGA+$U$+SOC electronic structure of NiRh$_2$O$_4$ in tetragonal phase. (c) The energy level positions for the spin-split and the crystal-field-split
  Ni $d$ and Rh $d$ states. For clarity, small splittings around $0.1$~eV are not marked in the figure.}
\end{figure}

The electronic structure of NiRh$_2$O$_4$, calculated within GGA+$U$ ($U_{\rm Ni}$=5 eV, $J_H$=1 eV) resulted in half-metallic solutions for both the high temperature cubic and the low temperature tetragonal phases. Calculations
show the spin splitting at Ni site to be large ($\approx \!\! 1$  eV) while that at Rh site is an order of magnitude smaller ($\approx \!\! 0.1$~eV), in accordance with the nominal magnetic and non-magnetic character of Ni$^{2+}$ and Rh$^{3+}$ respectively. 
In the high-symmetry cubic phase (see SM \cite{sm} for details), the octahedral 
crystal field around Rh splits the 4$d$ states into $t_{2g}$ and $e_{g}$ with a large splitting $\sim \!\! 3$  eV,
while the tetrahedral crystal field around Ni splits the 3$d$ states into $e$ and $t_2$ with a relatively smaller splitting $\approx \!\! 0.6$~eV. The $d$ states of high spin Ni are thus fully occupied in the up-spin channel; in the down-spin channel, the Ni $t_2$ states admixed with Rh $t_{2g}$ and O $p$ states cross the
Fermi level (E$_F$). The Rh $t_{2g}$ states are mostly occupied, except for the mixing with Ni states
in down spin channel, while Rh $e_g$ states are empty. This is in accordance with
nominal valence of Ni$^{2+}$ with 2 holes in $t_2$ manifold, and low-spin nominally $d^6$ occupancy of Rh.
This general picture remains valid also in the tetragonal phase as shown in Fig. 1.
The tetragonal distortion, however introduces additional splitting among the 
cubic symmetry split states. This splits the Ni $t_2$ states with Ni $d_{xy}$ level positioned
above Ni $d_{xz}/d_{yz}$ with splitting of $\approx$ 0.1  eV. One of the two holes of Ni thus occupies
the down spin $d_{xy}$ level, while the other hole occupies the down spin doubly degenerate $d_{xz}/d_{yz}$
levels. This leaves the GGA+$U$ solution half-metallic even in the tetragonal phase, as shown in
Fig.~1(a). The crystal and spin splittings at the tetragonal phase is shown
in Fig.~1(c), which further highlights the energetic proximity of Ni $t_2$ and Rh
$t_{2g}$ states in down-spin channel, driving the high degree of mixing between the two.
This mixing gives rise to a small nonzero magnetic moment $\approx$ 0.06-0.07 $\mu_B$ at the otherwise
nonmagnetic, low-spin, nominally $d^6$ Rh site, while the Ni moment
is found to be 1.5 -1.6$\mu_B$. The remaining moment lives on O sites, giving rise to a
net moment of 2 $\mu_B$/f.u in both cubic and tetragonal phases.

\begin{figure}[b]
\includegraphics[width=0.5\textwidth,keepaspectratio]{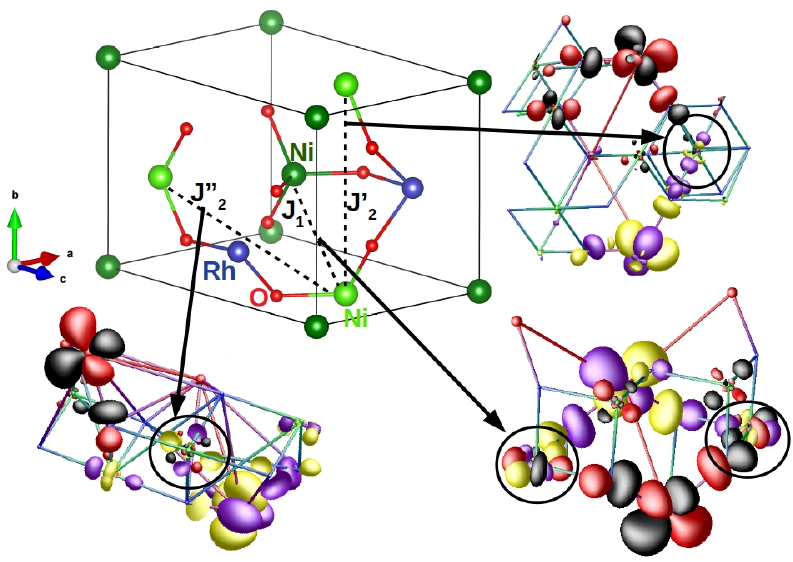}
\caption{(Color online) The exchange pathways for first neighbor ($J_1$) and inequivalent second neighbors ($J_2', J_2{''}$) in the low-temperature 
tetragonal phase of NiRh$_2$O$_4$. Shown are the overlap of effective Ni Wannier functions placed at NN, in-plane NNN and out of plane
NNN Ni sites, with circles indicating nonzero weight
at Rh sites in the pathway. Opposite sign parts of each Wannier function are colored differently.}
\end{figure}

Given the active orbital degrees of freedom at Ni site, we next explore the effect of SOC.
Within the GGA+$U$+SOC approach, the orbital state at Ni is derived from
the $d_{xz} \pm id_{yz}$ orbitals. Due to partial occupancy of both orbitals, Ni develops a large orbital
moment of $\sim\!\! 1.0 \mu_B$, supporting formation of a $S \!\!=\!\! 1$, $L\!\!=\!\! 1$ state. Repeating the calculation within 
GGA+SOC scheme, leads to a significantly smaller estimate of Ni orbital moment of $\approx$ 0.1 $\mu_B$, 
due to inability of GGA to capture the orbital polarization effect \cite{footnoteRh}.
While GGA+SOC splits the partially occupied
orbitally degenerate states in down spin channel, this splitting is insufficient to open an insulating
gap.
This situation is similar to that discussed
in case of FeCr$_2$S$_4$ \cite{Sarkar_PRB2009}.
The Coulomb correlation within GGA+$U$+SOC is thus crucial to produce a renormalized, large, orbital polarization  \cite{ani} which
drives the system insulating, with a $\sim\! 0.25$eV charge gap, as shown in Fig.~1(c).

We next estimate the Ni-Ni magnetic exchange from the knowledge of the effective hopping strengths and onsite
energies in the Wannier basis of Ni-$t_2$ only low-energy Hamiltonian (see SM for details). The dominant
 AFM interactions in cubic phase turn out to be between four nearest-neighbor (NN) Ni sites ($J_1$), which belong
to two different face-centered cubic (fcc) sublattices of the diamond lattice, and twelve next-nearest neighbor (NNN) 
Ni sites ($J_2$), which belong to the same fcc sublattice.
The tetragonal distortion splits the twelve NNN Ni-Ni interactions into four in-plane ($J^{_{'}}_2$) and eight out of
plane ($J^{_{''}}_2$) interactions (see Fig. 2). The substantial mixing between Ni and Rh states, makes the Ni-O-Rh-O-Ni superexchange 
paths strong, as seen from the overlap of Wannier functions in Fig. 2 (see encircled part).
The calculated exchanges are $J_1 \!\! \approx \!\! 1.2$~meV \cite{footnote1}, with $J^{_{'}}_2,J^{_{''}}_2 \!\! \approx\! 0.4 J_1$,
showing strong magnetic frustration.

{\it Single-site model. ---} Armed with the DFT results, we construct an effective single-site Hamiltonian 
for the $L\!\!=\!\!1$ and $S\!\!=\!\! 1$ state, taking into account the tetragonal distortion ($\delta$) and SOC ($\lambda$);
\be
H = - \delta L_z^2 + \lambda \vec L \cdot \vec S
\ee
Based on DFT inputs, we consider the limit $\delta \gg \lambda$, and show that this leads to a 
simple, yet complete, understanding of the low temperature phenomenology of this distorted spinel.

In the regime $\delta \gg \lambda$, we start by constructing
orbital eigenstates with well-defined $L_z$, which leads to a ground doublet with $L_z = \pm 1$ and an excited orbital singlet with $L_z=0$
which is split off by an energy $\delta$.

Next, let us take the spin degrees of freedom into account, which couple via SOC $\lambda \ll \delta$. 
The dominant SOC coupling is $\lambda L_z S_z$, which leads to a sequence
of states in increasing order of energy which we label by $|L_z,S_z\ra$:
\bea
E^0_0~[2] \!\!&=&\!\! -\delta - \lambda: |\pm,\mp\ra;~~ E^0_1~[2] = -\delta: |\pm,0 \ra \nonumber \\
E^0_2~[2] \!\!&=&\!\! -\delta+ \lambda: |\pm,\pm \ra;~~ E^0_3~[3] = 0: |0,0\ra, |0,\pm\ra
\eea
with degeneracies shown in square brackets.
We can perturbatively treat $\lambda (L^+ S^- + L^- S^+)/2$, since it only couples the low lying 
states at $E^0_{0,1,2}$ to the
high energy states at $E^0_3$. Let us define the symmetric state $|e \ra = (|+,-\ra + |-,+\ra)/\sqrt{2}$.
We then find the sequence of states, with energies defined relative to the ground state,
\bea
\!\!\!\! \Delta_0 =0 &:&
|\psi_0 \ra \approx |e \ra - \sqrt{2} \frac{\lambda}{\delta}|0,\!0\ra\\
\Delta_1 \! \approx \! 2 \frac{\lambda^2}{\delta} &:&
|\psi_1 \ra = \frac{|+,-\ra - |-,+\ra}{\sqrt{2}}\\
\Delta_2 \!\approx\! \lambda + \frac{\lambda^2}{\delta} &:& |\psi_{2,\pm} \ra \approx |\pm,0\ra - \frac{\lambda}{\delta} |0,\pm\ra \\
\Delta_3 \approx 2 \lambda + 2\frac{\lambda^2}{\delta} &:& |\psi_{3,\pm} \ra = |\pm,\pm\ra \\
\Delta_4 \approx \delta + \lambda + 3 \frac{\lambda^2}{\delta} &:& |\psi_{4\pm}\ra \approx |0,\pm\ra + \frac{\lambda}{\delta} |\pm,0\ra\\
\Delta_5 \approx \delta + \lambda + 4 \frac{\lambda^2}{\delta} &:& |\psi_5 \ra \approx |0,0\ra + \frac{\sqrt{2}\lambda}{\delta} |e \ra
\eea
With these states and energies in hand, and a choice $\lambda \!\!\sim\!\! 10$meV and $\delta \!\!\sim \!\! 100$meV,
we readily obtain a broad-brush understanding of 
some key experimental observations as summarized below. (The choice of $\delta \!\!\sim \!\! 100$meV agrees with the
spin-averaged crystal field splitting between $d_{xy}$ and $d_{yz}/d_{xz}$ orbitals from our DFT).
We present further arguments against alternative scenarios in the SM \cite{sm}.


{\it Ground state:}
We find that the ground state is a non-magnetic singlet. This is consistent with the lack of any magnetic order down to the lowest
temperature in this material \cite{footnote2}. In contrast to previous proposals of non-magnetic $S_z=0$ state, our proposed
state is a spin-orbit entangled ``Schrodinger-cat'' type state arising from weak off-diagonal SOC induced splitting of a doublet.

{\it Thermodynamics:} Since the gap to the states $|\psi_{4\pm}\ra,|\psi_5\ra$ are large, we expect to recover only an entropy $S_{\rm low} = R\ln 6$ for $T < 300$K,
consistent with specific heat measurements \cite{Chamorro_PRM2018}
carried out up to room temperature (which corresponds to $T \ll \Delta_4$). At low temperatures, the state at $\Delta_1$ leads to a
Schottky peak in $C/T$ at $T \sim 10$K from the level $|\psi_1\ra$ (see SM \cite{sm}). It is not clear why this peak has not been observed; one possibility is that it
may be affected by defects, which also likely lead to the observed spin freezing for $T \lesssim 6$K. 
The higher levels $|\psi_{2\pm}\ra$ lead to a broad Schottky anomaly for $T \sim 30$-$40$K, similar to the experiments.

{\it Neutron scattering:} 
Our results for the local dynamical spin correlation function $S_{\rm loc}(\omega)$ are summarized in
Fig.~3(a).
The first excited state is nondegenerate, separated by an energy $\Delta_1 \approx 2\lambda^2/\Delta \! \approx \! 2$meV. 
We note that $|\psi_0\ra$ and $|\psi_1\ra$ are connected via $S^z$, so $|\psi_1\ra$ should be visible in non-spin-flip scattering, but
appears difficult to observe due to the resolution and the background, as well as possibly defects.
The second excited state is a doublet $|\psi_{2,\pm} \ra$ with an energy gap $\Delta_2 \! \approx\! \lambda \!+\! \lambda^2/\Delta$.
We propose that it is
this doublet state which has been observed as a gapped mode 
in INS experiments \cite{Chamorro_PRM2018}. The above parameter choice leads to the gap $\Delta_2 \approx 11$meV, in crude agreement with the data.
Based on our analysis, the states
$|\psi_{3\pm} \ra$ at an energy gap $\Delta_3 \approx 22$meV and the singlet state $|\psi_5\ra$ at a gap $\Delta_4\approx 108$meV are both 
``dark states'', invisible to neutrons due to vanishing matrix elements. Finally, $|\psi_{4\pm}\ra$ with a gap $\Delta_5 \approx 107$meV 
should be visible but with spectral weight much smaller that of $|\psi_{2\pm}\ra$. This is a prediction for future INS experiments.

{\it Magnetic susceptibility:}
The numerically computed single-site magnetic susceptibility $\chi$ can be fitted to an apparent ``Curie-Weiss'' form 
$\chi(T) = \chi_0 + \alpha/(T - T_0)$,
 with a negligible background $\chi_0 \sim 10^{-5}$, an effective ``Curie-Weiss'' scale $T_0 \approx 16(2)$K, and $\alpha\approx 0.85(2)$ (see SM) \cite{footnote3}. 
 In analyzing experiments, we expect
$\chi_0$ will get lumped together with a background van Vleck type contribution which is conventionally subtracted. Our
estimate for $T_0$ is small and ``ferromagnetic'' in sign, so that the $T_0^{\rm expt} \approx -11$K observed in experiments \cite{Chamorro_PRM2018}
must be attributed to weak residual intersite AFM exchanges on the scale of $\sim \!\! 1$ meV.
Setting the fitted value of $\alpha \!\! \equiv\!\! S_{\rm eff} (S_{\rm eff}+1)/3$, 
yields an effective spin $S_{\rm eff}\!\!=\!\!1.4$ (or an effective magnetic moment $p_{\rm eff} \!\! \sim \!\! 3.6 \mu_B$), larger
than a spin-only value $S\!\!=\!\! 1$ as in experiments \cite{Chamorro_PRM2018}.

{\it Inter-site exchange. ---}
We next incorporate inter-site interactions via a simple $J_1$-$J_2$ Heisenberg exchange model
$H_{\rm ex} \!\!=\!\! \frac{1}{2} \sum_{i,j} J_{ij} {\bf S}_i \cdot {\bf S}_j$. In order to compute the
spin dynamics in the low energy Hilbert space, we
introduce, in the spirit of slave-boson theory \cite{Sachdev_PRB1990,GangChen_J0_2018},
four local boson operators, $c^\dg_0, c^\dg_1, d^\dg_{\pm}$, which respectively create states $|\psi_0\ra$, $|\psi_1\ra$, 
and $|\psi_{2\pm}\ra$. Projecting the Heisenberg model to this
Hilbert subspace, and imposing the local completeness constraint $c^\dg_0 c^\pdg_0 \!+\! c^\dg_1 c^\pdg_1 \!+\! d^\dg_\alpha d^\pdg_\alpha \!=\! 1$
(with an implicit sum on $\alpha=\pm$), we find that the site spin-$1$ operators may be approximated as
$S^z = (c^\dg_1 c^\pdg_0 + c^\dg_0 c^\pdg_1)$ and
$S^\pm = (c_0^\dg\pm c^\dg_1) d^\pdg_\pm + d^\dg_{\mp} (c^\pdg_0 \mp c^\pdg_1)$.
At mean field level, we replace $c_0 \!\to\!  \la c_0\ra $, 
and retain leading powers in $\la c_0\ra$, to arrive at 
the Hamiltonian $H_{\rm tot} \!=\! H_{\rm loc} \!+\! H_{\rm ex} \!+\! H_{\rm con}$,
where
\bea
H_{\rm loc} &=& \sum_i (\Delta^\pdg_1 c^\dg_{i 1} c^\pdg_{i 1} + \Delta^\pdg_2 d^\dg_{i \alpha} d^\pdg_{i \alpha} ) \\
H_{\rm ex} &=& \frac{1}{4} \la c_0 \ra^2 \sum_{i,j} J_{ij} \left[ (d^\dg_{i \alpha} d^\pdg_{j \alpha} + d^\dg_{i \alpha} d^\dg_{j \bar{\alpha}} + h.c.) \right. \nonumber \\
&+& \left. 2 (c^\dg_{i 1} + c^\pdg_{i 1})(c^\dg_{j 1} + c^\pdg_{j 1}) \right]\\
H_{\rm con} &=& -\mu \sum_i (c^\dg_{i 1} c^\pdg_{i 1} + d^\dg_{i \alpha} d^\pdg_{i \alpha} + \la c_0 \ra^2 -1)
\eea
The different pieces correspond respectively to the local single-site Hamiltonian, the inter-site exchange Hamiltonian, and the constraint imposed 
(on average) via the Lagrange multiplier $\mu$. We note that the $c$ and $d$ bosons are decoupled at this order (except for the constraint).
We can thus solve this in momentum space separately for these two sectors, leading to
\bea
H_{\rm tot} &=& \sum_{\bk,\sigma} (E^\sigma_\bk \alpha^\dg_{\bk,\sigma}  \alpha^\pdg_{\bk,\sigma} + 
\tilde{E}^\sigma_\bk \beta^\dg_{\bk,\sigma}  \beta^\pdg_{\bk,\sigma}) - 2\mu \sum_\bk \la c_0 \ra^2 \nonumber \\
&+& \sum_{\bk\sigma} (\frac{1}{2}  E^\sigma_\bk + \tilde{E}^\sigma_\bk)  - \sum_\bk (\Delta_1 + 2 \Delta_2 - 5\mu)
\eea
Here, $\sigma=\pm$, and the excitation energies are given by
\bea
\!\!\!\!\!\!\!\! E^{\sigma}_\bk \!\!&=&\!\! (\Delta_1 \!-\! \mu)^{1/2} [\Delta_1 \!- \! \mu \!+\! 2 \la c_0 \ra^2 (\sigma J_1 |\gamma_\bk| \!+\! J_2 \eta_\bk)]^{1/2}\\
\!\!\!\!\!\!\!\! \tilde{E}^{\sigma}_\bk \!\!&=&\!\! (\Delta_2 \!-\! \mu)^{1/2} [\Delta_2 \!- \! \mu \!+\! \la c_0 \ra^2 (\sigma J_1 |\gamma_\bk| \!+\! J_2 \eta_\bk)]^{1/2}
\eea
with $\gamma^\pdg_\bk= \sum_{\ell_1} e^{i \bk\cdot \ell_1}$ and $\eta^\pdg_\bk= \sum_{\ell_2} e^{i \bk\cdot \ell_2}$, 
where $\ell_1, \ell_2$ are respectively the $4$ nearest-neighbor and $12$ next-neighbor vectors.
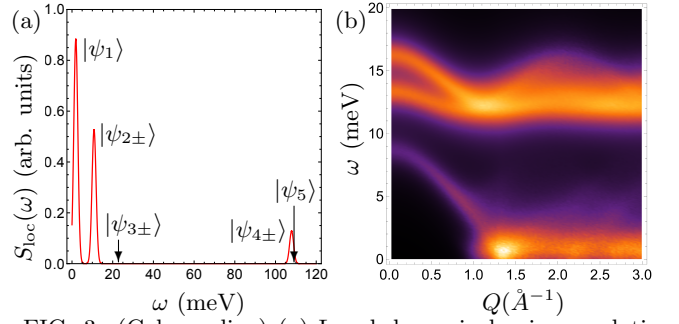
\begin{figure}[tb]
\begin{overpic}[width=0.21\textwidth]{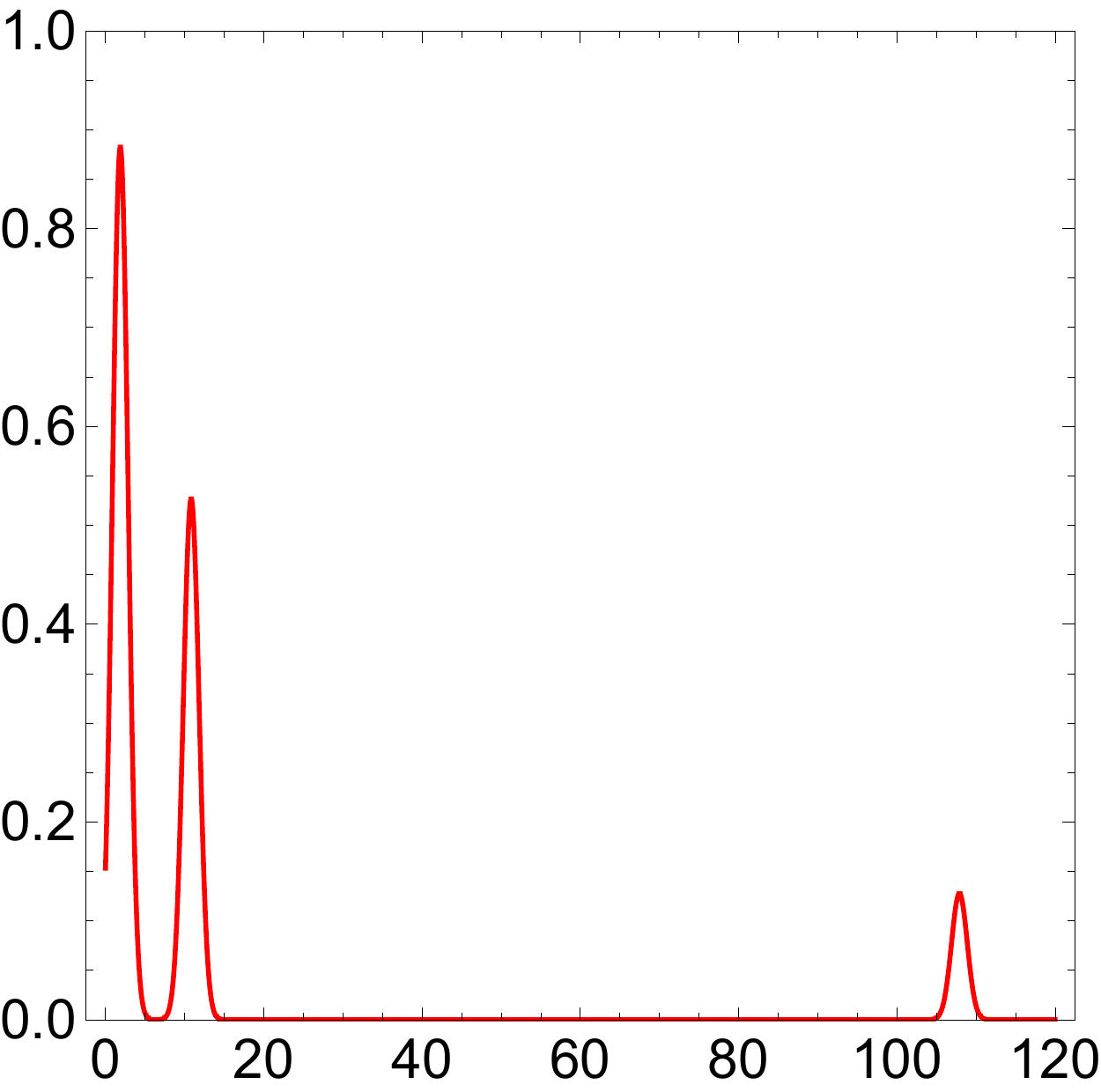}\put(38,-10){$\omega$ (meV)}\put(-11,7){\rotatebox{90}{$S_{\rm loc}(\omega)$~(arb. units)}}
\put(13,80){$|\psi_1\ra$}
\put(19,50){$|\psi_{2\pm}\ra$}
\put(21,18){$|\psi_{3\pm}\ra$}
\put(26,15){\color{black}\vector(0,-1){8}}
\put(65,16){$|\psi_{4\pm}\ra$}
\put(81,30){$|\psi_{5}\ra$}
\put(88,27){\color{black}\vector(0,-1){20}}
\put(-12,90){(a)}
\end{overpic}~~~~
\begin{overpic}[width=0.21\textwidth]{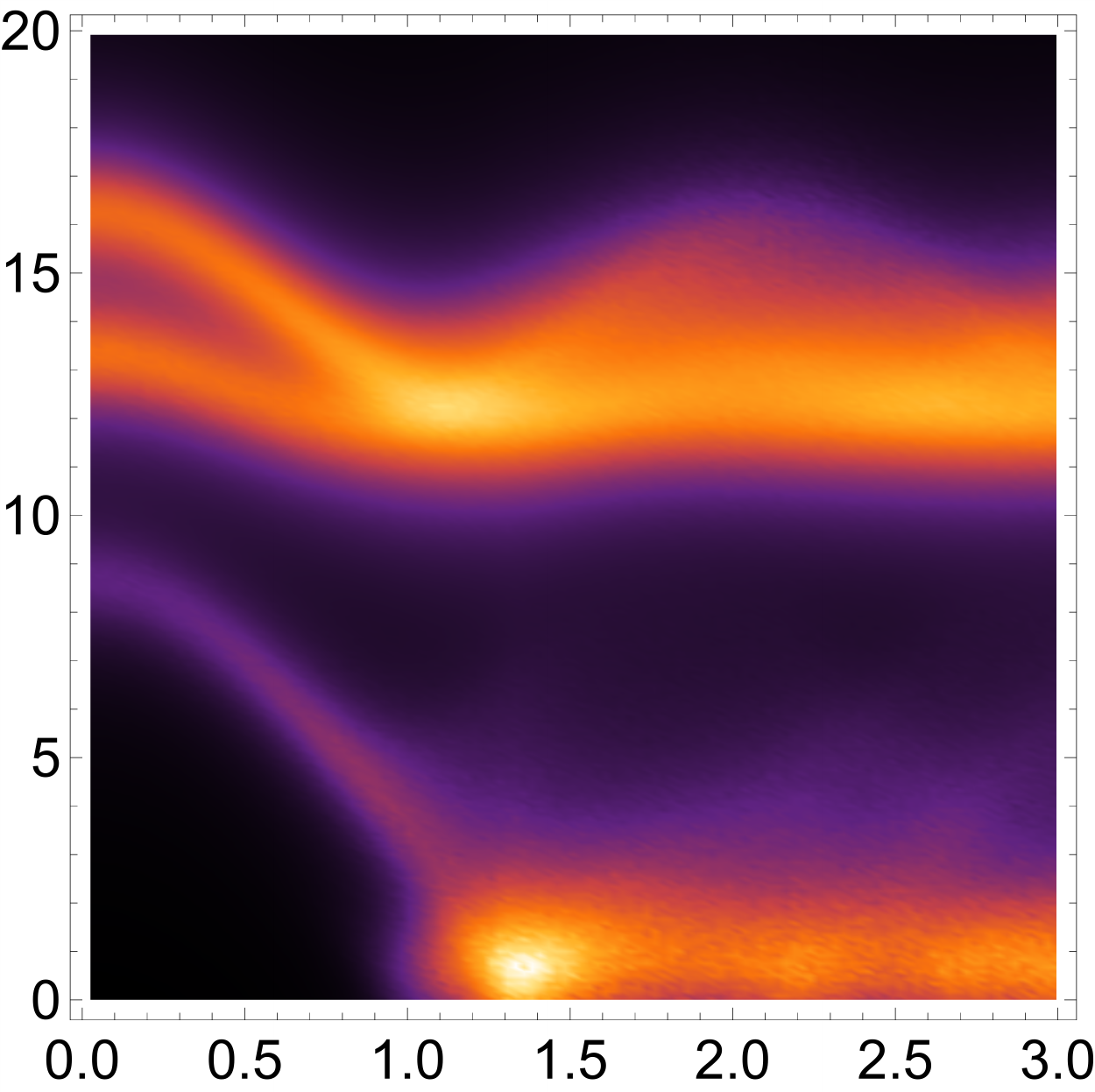}\put(40,-10){$Q (\AA^{-1})$}\put(-12,90){(b)}\put(-10,38){\rotatebox{90}{$\omega$~(meV)}}
\end{overpic}
\caption{(Color online) (a) Local dynamical spin correlation function $S_{\rm loc}(\omega)$ within single-site model. Peaks are labelled by
relevant excited states, arrows indicate ``dark states'' invisible to neutrons due to vanishing matrix elements.
(b) Intensity plot (arbitrary units) of powder-averaged INS spin structure factor incorporating inter-site exchange,
${\cal S}(Q,\omega)$, as a function of wavevector $Q$
and energy $\omega$ (with $1$meV broadening to mimic experimental resolution); see text for details.}
\label{fig:sqw}
\end{figure}
We choose $\Delta_1 \!\!=\!\! 1.8$meV and $\Delta_2 \!\!=\!\! 11$meV based on the single-site model, and
$J_1 \!\!=\!\! 1.2$meV and $J_2/J_1 \!\!=\!\! 0.4$ from our DFT.
Using these parameters,  we minimize the ground state energy with respect to $\la c_0 \ra^2$
while choosing $\mu$ to satisfy the constraint.
We find the optimal $\la c_0 \ra^2 \!\approx\! 0.7$ and $\mu \!\approx\! -2.1$meV.

The resulting weighted and powder-averaged dynamic spin structure factor relevant to INS experiments,
${\cal S}(\bQ,\omega)=\sum_\alpha (1-Q^2_\alpha/Q^2) S_{\alpha\alpha}(\bQ,\omega)$, including a $1$meV broadening to 
mimic the experimental resolution but ignoring form factors,  
is plotted in Fig.~\ref{fig:sqw}(b) (see also SM \cite{sm}). The upper gapped mode, arising from the $|\psi_{2\pm}\ra$ states, is in rough agreement 
with INS observations of a gapped dispersive mode \cite{Chamorro_PRM2018}; we find that it really consists of two peaks due to two sublattices 
on the diamond lattice.
The lower gapped mode is the ``optical branch'' of the $|\psi_{1}\ra$ state. It collapses in energy, with increasing $Q$, from $\sim\!\! 8$meV down to
$\sim \!\! 0.5$meV, and persists as an intense small-gap band, robust against magnetic condensate formation due to frustrating
$J_2$ exchange. The lower energy ``acoustic branch'' of the $|\psi_1\ra$ state is also gapped, but it has negligible intensity and is not 
visible here (see SM \cite{sm}). The small-$Q$ behavior
depicted here may be partly masked by neutron kinematic constraints.

{\it Summary and discussion. ---} We have combined DFT and model calculations to address the mystery of NiRh$_2$O$_4$, broadly capturing the
existing thermodynamic and INS observations. In light of our work, it may be useful to revisit the low temperature specific heat and low
energy INS on higher purity samples, and use INS to probe the predicted high energy crystal field level around $\sim 110$~meV. THz spectroscopy
\cite{Armitage_PRL2015,Armitage_PRX2018}
on NiRh$_2$O$_4$ could help to test our prediction of the ``optical'' $|\psi_1\ra$ mode at $Q=0$, and infrared spectrocopy could 
be used to measure the insulating charge gap. It may be possible to use resonant inelastic
X-ray scattering at a Ni-edge \cite{Keimer_PRX2018} to look for the predicted $|\psi_{3\pm}\ra$ and $|\psi_5\ra$ ``dark states'' which are
invisible to neutrons. Finally, our work suggests that NiRh$_2$O$_4$ does not realize a topological quantum paramagnet. However, it guides future searches by
suggesting that tetragonal compression, presumably achievable by application of uniaxial strain, may provide the means to quench orbital angular momentum and suppress
SOC effects, potentially stabilizing more exotic phases.

AP acknowledges support through a Discovery Grant from NSERC of Canada. 
TS-D thanks the Department of Science and Technology, India for the support through a Thematic Unit of Excellence.


\begin{thebibliography}{48}
\bibitem[{foo({\natexlab{a}})}]{footnoteRh}
\bibinfo{note}{The calculated orbital moment at Rh site is found to be smaller
  than that of Ni ($\approx$ 0.05 $\mu_B$) both within GGA+SOC and GGA+$U$+SOC,
  suggesting near quenching of orbital moment at Rh site.}
\bibitem[{ani()}]{ani}
\bibinfo{note}{The calculated spin-ion anisotropy is large ($\approx 40$~meV)
  with easy axis pointing along $(111)$ direction}.
\bibitem[{foo({\natexlab{b}})}]{footnote1}
\bibinfo{note}{Varying the $U_{\rm Ni}$ value between $5$-$7$ eV, the dominant
  interaction $J_1$ is found to vary between $1.2$-$0.8$ meV, with little
  changes in the ratios of magnetic exchanges.}
\end{thebibliography}

\end{document}